# From virtual work principle to least action principle for stochastic dynamics

Qiuping A. Wang

Institut Supérieur des Matériaux et Mécaniques Avancés du Mans, 44 Av. Bartholdi,

72000 Le Mans, France

Abstract

After the justification of the maximum entropy principle for equilibrium mechanical system from the principle of virtual work, i.e., the virtual work of microscopic forces on the elements of a mechanical system vanishes in thermodynamic equilibrium, we present in this paper an application of the same principle to dynamical systems out of equilibrium. The aim of this work is to justify a least action principle and the concurrent maximum path entropy principle for nonequilibrium thermodynamic systems.

## 1) Introduction

The least action principle[1] (LAP) first developed by Maupertuis[1][2] is originally formulated for regular dynamics of mechanical system. One naturally ask the question about the destiny of the principle when the system is subject to noise so that the dynamics becomes irregular and stochastic such as in a diffusion. There have been many efforts to answer this question. One can count Onsager[3] and De Broglie[4] among the first scientists who were interested in developing least action principle or similar approach for random dynamics. Other efforts have also been made in the fields such as random dynamics[5][6], stochastic mechanics[7][8], quantum theory[9] and quantum gravity theory[10]. A common feature of these works is to mimic the mathematical formalism of LAP using either the original Lagrange action or some different effective action, but without considering explicitly the role of dynamical uncertainty in the optimization calculus. For example, we sometimes see expression such as $\delta \overline{R} = \overline{\delta R}$ concerning the variation of a random variable $R$ with expectation $\overline{R}$, where the variation of uncertainty due to the variation of $R$ and of probability distribution is neglected, which is of course not true in general.

To amend this incompleteness of optimization, an extension of the Maupertuis principle using the Lagrange action (see definition below) was suggested for describing stochastic motion of mechanical system subject to noise[11][12]. A new ingredient is the introduction of informational entropy or probabilistic uncertainty in variational calculus. This approach led to a so called stochastic action principle (SAP) given by

$$\overline{\delta A} = 0 \qquad (1)$$

where $A$ is the Lagrange action and the $\overline{\delta A}$ its variation averaged over all the possible paths between two points $a$ and $b$ in position space. When the noise is vanishing, Eq.(1) becomes the usual principle $\delta A = 0$. In fact, Eq.(1) is equivalent to a maximization of path entropy $S_{ab}$ defined by

$$\delta S_{ab} = \eta \left( \delta \overline{A} - \overline{\delta A} \right) \qquad (2)$$

where $\eta$ is a characteristic constant of the dynamics[11].

---

[1] We continue to use the term "least action principle" here considering its popularity in the scientific community. We know nowadays that the term "optimal action" is more suitable because the action of a mechanical system can have a maximum, or a minimum, or a stationary for real paths[18].

This formalism seems useful from several points of view. It has a diffusion probability in exponential of action $\propto e^{-\eta A}$ if the path entropy is of Shannon form[11], or, equivalently, if the distribution of the perturbed system in position space is of Gaussian type such as the numerical results obtain in [12]. For free diffusing particles, this is the transition probability of Brownian motion. For particles in arbitrary potential energy, a Fokker-Planck equation for normal diffusion can be derived from this diffusion probability[11].

This approach has many underlying basic assumptions which are not always obvious from physical point of view. Following questions can be asked. 1) Why one should use the Lagrange action instead of other ones already used in other formulation of stochastic dynamics? 2) Why the variation $\overline{\delta A} = (\overline{\delta A} - \overline{\delta S_{ab}}/\eta) = 0$ instead of $\overline{\delta A} = 0$ should be considered for the perturbed mechanical system? 3) Why $S_{ab}$, as an information or uncertainty measure, must be at maximum for equilibrium state or some special states out of equilibrium? And, finally, why $S_{ab}$ may take the Shannon form?

The last two questions are in fact related to the longstanding questions around MEP since the appearance of MEP by Jaynes for inference theory[13], i.e., the original version based on the uniqueness of Shannon entropy as the maximizable information measure. Although MEP, as a variational method, is actually almost a doctrine for many and used often for equilibrium as well as for nonequilibrium system, the justification and the validity of MEP for inference theory and physics are still subject to considerable criticism and controversy[13][14][15].

In the present work, we try to find answers to the above questions for thermodynamic system from the viewpoint of mechanics. Eq.(1) will be derived from a more obvious and widely accepted principle of physics: the principle of virtual work[16][17]. This latter is a simple, palpable, successfully used principle in analytical mechanics theory and mechanical engineering. One can find its origin in a very simple fact that a point is in static equilibrium under the action of different forces who cancel themselves, and that a body is in static equilibrium when its potential energy is at minimum. It is expected that the link to such a basic principle will make MEP (as well as SAP) less obscure and mysterious as it appears with so many polemics and controversies around it. In what follows, we look at mechanical systems out of equilibrium. The term "entropy" is used as a measure of uncertainty or randomness of stochastic motion. It will be indicated if "entropy" is used in the sense of equilibrium thermodynamics.

2) Principle of least action

The least action principle is well formulated for non-dissipative Hamiltonian system satisfying following equations [2]:

$$\dot{x}_k = \frac{\partial H}{\partial P_k} \text{ and } \dot{P}_k = -\frac{\partial H}{\partial x_k} \text{ with } k=1,2, \ldots g \qquad (3)$$

where $x_k$ is the coordinates, $P_k$ the momentum, $H$ the Hamiltonian given by $H = T + V$, $T$ the kinetic energy, $V$ the potential energy, and $g$ the number of degrees of freedom of the system. The Lagrangian is defined by $L = T - V$.

The least action principle stipulates that the action of a motion between two point $a$ and $b$ in the configuration space defined by the time integral $A = \int_a^b L dt$ on a given path from $a$ to $b$ must be a stationary on the unique true path for given period of time $\tau$ of the motion between the two points, i.e.,

$$\delta A|_\tau = 0 \qquad (4)$$

In what follows, we will drop the index $\tau$ of the variation and the action variation is always calculated for fixed period of time $\tau$. This principle yield the famous Lagrange-Euler equation given by

$$\frac{\partial}{\partial t}\frac{\partial L}{\partial \dot{x}_k} - \frac{\partial L}{\partial x_k} = 0 \qquad (5)$$

With $\dot{x}_k = \frac{\partial x_k}{\partial t}$. These above equations underlie a completely deterministic dynamic process: if the time period of the motion is given, there is only one path between two given points so that all the states of the systems are completely determined by Eq.(5) for every moment of the motion. However, this deterministic character of the dynamics does not exist any more when the motion becomes random and stochastic[11]. This is the physical situation we encounter in the case of thermodynamic systems either in equilibrium or out of equilibrium.

3) Principle of virtual work

In mechanics, a virtual displacement of a system is a kind of hypothetical infinitesimal displacement with no time passage and no influence on the forces. It should be perpendicular to the constraint forces. The principle of virtual work says that the total work done by all forces acting on a system in static equilibrium is zero for any possible virtual displacement. Let us suppose a simple case of a system of $N$ points of mass in equilibrium under the action of $N$ forces $\boldsymbol{F}_i$ ($i=1,2,…N$) with $\boldsymbol{F}_i$ on the point $i$, and imagine virtual displacement of each point $\delta \vec{r}_i$ for the point $i$. According to the principle, the virtual work $\delta W$ of all the forces $\boldsymbol{F}_i$ on all $\delta \vec{r}_i$ vanishes for static equilibrium, i.e.

$$\delta W = \sum_{i=1}^{N} \vec{F}_i \cdot \delta \vec{r}_i = 0 \tag{6}$$

This principle for static equilibrium problem was extended to "dynamical equilibrium" by d'Alembert[17] who added the initial force $-m_i \vec{a}_i$ on each point of the system in motion

$$\delta W = \sum_{i=1}^{N} (\vec{F}_i - m_i \vec{a}_i) \cdot \delta \vec{r}_i = 0 \tag{7}$$

where $m_i$ is the mass of the point $i$ and $\vec{a}_i$ its acceleration. From this principle, we can not only derive Newtonian equation of dynamics, but also other fundamental principles such as least action principle. This principle has been used to give a derivation of MEP for equilibrium system[19] which we recapitulate as follows.

4) Maximum entropy for equilibrium system

Due to the randomness of the dynamics, the deterministic character of Eq.(7) must be changed in order to introduce the probabilistic description of the system in random motion. This is to be done by using the notion of statistical ensemble. Suppose $\delta W_j$ is the virtual work of all the forces acting on every element of a given system at a microstate $j$ in a canonical ensemble, it can be proven that[19]

$$\delta W_j = \left( -\sum_{i=1}^{N} \delta e_i \right)_j = -\delta E_j \tag{8}$$

which is the variation of energy $E_j$ of a system of the ensemble at microstate $j$ due to the virtual work. Hence the average virtual work for the whole ensemble is given by

$$\delta W = \sum_{j=0}^{w} p_j \delta W_j = -\sum_{j=0}^{w} p_j \delta E_j = -\overline{\delta E} \tag{9}$$

where $p_j=p(E_j)$ is the probability that the system is found at the state $j$. Since we have $\delta \sum_{j=0}^{w} p_j E_j = \sum_{j=0}^{w} p_j \delta E_j + \sum_{j=0}^{w} E_j \delta p_j$, Eq.(9) can still be changed into $\overline{\delta E} = \delta \overline{E} - \sum_{j=0}^{w} E_j \delta p_j$ where $\overline{E} = \sum_{j=0}^{w} p_j E_j$ is the usual internal energy. This relationship can be seen as a virtual version of the first law of thermodynamics if we identify $\sum_{j=0}^{w} E_j \delta p_j$ to the heat transfer, i.e.,

$$\overline{\delta E} = \delta \overline{E} - \delta Q. \qquad (10)$$

where $\delta Q = \sum_{j=0}^{w} E_j \delta p_j$ is a well known relationship derived within the usual Boltzmann-Gibbs statistical mechanics with Shannon entropy and exponential $p_j$. Here it is derived from the microscopic virtual works on each particle with only one constraint: the first law of thermodynamics or the conservation of energy. No hypothesis is considered about the probability and entropy property. If we further suppose a reversible virtual process, we can use the second law to write $\delta S = \beta \delta Q = \beta \sum_{j=0}^{w} E_j \delta p_j$ where $S$ is the thermodynamic entropy. Application of the principle of virtual work Eq.(7) to Eq.(10) yields

$$\delta W = \delta(\frac{S}{\beta} - \overline{E}) = 0 \qquad (11)$$

This equation must be considered as the condition of the *dynamical equilibrium* of the canonical ensemble. It is an optimization of the functional $(S - \beta \overline{E})$ for thermodynamic equilibrium. In other words, for a random dynamics to be in equilibrium, the difference between the heat (or entropy as a measure of disorder) and internal energy must be optimized. This variational method is to be used with the constraint associated with the normalization $\sum_{j=0}^{w} p_j = 1$ or $\sum_{j=0}^{w} \delta p_j = 0$, i.e.,

$$\delta(S + \alpha \sum_{j=0}^{w} p_j - \beta \sum_{j=0}^{w} p_j E_j) = 0 \qquad (12)$$

which is nothing but the variational approach MEP of Jaynes. However, it should be noticed that in the variational method of Eq.(11) or (12), there is no restriction on the functional form of entropy $S$, which is an essential difference between the MEP by virtual work principle and its original version by Jaynes who argued for the use of Shannon entropy in MEP from the inferential point of view and based his arguments on the subjective character of the probability notion. For Jaynes, MEP is only an mathematical principle without physics in it[13][15]. But in the present framework, 1) MEP is a law of physics since it can be derived or justified from a most fundamental physics principle, and 2) the entropy in this MEP can take in principle whatever form if any for equilibrium system. *S* is of Shannon form if and only if the probability distribution of energy is exponential as in the Boltzmann-Gibbs statistics[19][20].

We would like to stress that the above conclusion is only valid for the ensemble of equilibrium system and that *S* must be the entropy the second law of thermodynamics since the second laws with reversible virtual process has been considered in the derivation. However, the mathematical formalism itself is not restricted to equilibrium ensemble. The reason for this is in Eq.(9), a natural consequence of Eq.(8) for the virtual work calculated from microscopic consideration. As discussed above, Eq.(9) can be written as

$$\delta W = -\delta \overline{E} + \delta \Omega \qquad (13)$$

here $\delta \Omega = \sum_{j=0}^{w} E_j \delta p_j$ is not necessarily the heat transfer $\delta Q$ if the system is not in equilibrium. Applying the principle of virtual work of d'Alembert to Eq.(13), we obtain

$$\delta(\Omega - \overline{E}) = 0. \qquad (14)$$

This is the optimization of the quantity $(\Omega - \overline{E})$ for any system at any moment whether or not it is in equilibrium, since the principle of virtual work of d'Alembert does apply for a moving system at any moment. For nonequilibrium system, obviously one cannot talk about variation of $\Omega$ in connection with the thermodynamic entropy *S* or heat transfer $\delta Q$. A detailed discussion of this approach needs careful definition of an entropy or information as a measure of nonequilibrium disorder in taking into account eventual heat transfer $\delta Q$, which will be reported in another paper.

In what follows, we still consider a statistical ensemble of mechanical systems out of equilibrium. But unlike the above treatment of equilibrium system where the virtual work principle was used for a given moment of the evolution, we will consider the trajectories of the ensemble in the position space. The virtual work will be calculated for an ensemble of points on the trajectories.

5) Stochastic least action principle

We have an ensemble of Hamiltonian systems (without energy dissipation) out of equilibrium. A system is composed of $N$ particles moving in the $3N$ dimensional position space starting from a point $a$. If the motion was regular, all the systems in the ensemble would follow a single $3N$-dimensional trajectory from $a$ to a given point $b$ according to the least action principle. But due to the random motion of the particles, every system in the ensemble is subject to irregular motion with some fluctuation as if there were random forces perturbing the systems. In this case, the systems can take different paths from $a$ to $b$ as shown in [11] and [12].

Now let us look at the random dynamics of a single system following a trajectory, say, $k$, from $a$ to $b$. At a given time $t$, the total force on a particle $i$ in the system is denoted by $\vec{F}_i(t)$ and the acceleration by $\vec{a}_i(t)$ with an inertial force $-m_i\vec{a}_i(t)$ where $m_i$ is its mass. The virtual work at this moment on a virtual displacement $\vec{\delta r}_{ik}$ of the particle $i$ on the trajectory $k$ should be

$$\delta W_i(t) = [\vec{F}_i(t) - m_i \vec{a}_i(t)]_k \cdot \vec{\delta r}_i \qquad (15)$$

Summing this work over all the particles, we obtain

$$\delta W_k(t) = \sum_{i=1}^{N} (\vec{F}_i - m_i \vec{a}_i)_k \cdot \vec{\delta r}_{ik} \qquad (16)$$

Remember that the principle of virtual work of d'Alembert can readily be applied at this moment as the principle is valid for any moment of a motion. But taking into account the fact that the system is in evolution on a trajectory, we continue the calculation of virtual work on that trajectory over which the system travels during the period $\tau$. The virtual work in Eq.(16) can be calculated for any moment over $\tau$ or any point over the trajectory $k$. Thus one has a series of equations like Eq.(16) for a finite number of points arbitrarily close one to another over the whole trajectory. For an infinitesimal time interval from $t$ to $t+dt$ in which the force and acceleration on each particle do not change, the virtual work during $dt$ at time $t$ on a small segment of $k$ must be proportional to $\delta W_k(t)dt$. Thus the total virtual work over the trajectory $k$ can be roughly given by

$$\Delta W_k \propto \int_a^b \delta W_k(t)\,dt = \int_a^b \sum_i [\vec{F}_i(t) - m_i \vec{a}_i(t)]_k \cdot \delta \vec{r}_{ik}\,dt. \qquad (17)$$

Now we have to taken into account the fact that this considered system is only one of a large number of systems of a statistical ensemble, all traveling from *a* to *b* during $\tau$ following different paths. So for the ensemble, the total virtual work is the statistical average of the virtual works for every system. Without loss of generality, we consider discrete paths denoted by $k=1,2 \ldots w$ (if the variation of path is continuous, the sum over $k$ must be replaced by path integral between $a$ and $b$[9]). Suppose $p_k$ is the probability that the path $k$ is taken by the systems from $a$ to $b$, the average virtual work is given by

$$\Delta W = \sum_{k=1}^{w} p_k \Delta W_k \propto \sum_{k=1}^{w} p_k \int_a^b \sum_{i=1}^{N} (\vec{F}_i - m_i \vec{a}_i)_k \cdot \delta \vec{r}_{ik}\,dt. \qquad (18)$$

In what follows, for the sake of simplicity, we consider only one degree of freedom in Eq.(18), say, $x$. It follows that

$$\Delta W \propto \sum_{k=1}^{w} p_k \int_a^b \sum_{i=1}^{N} (F_{xi} - m\ddot{x})_k\, \delta x_{ik}\,dt = \sum_{k=1}^{w} p_j \int_a^b \sum_{i=1}^{N} (-\frac{\partial H_i}{\partial x_i} - \dot{p}_{xi})_k\, \delta x_{ik}\,dt \qquad (19)$$

$$= \sum_{k=1}^{w} p_k \sum_{i=1}^{N} \int_a^b (\frac{\partial L}{\partial x_i}\delta x_i + (\frac{\partial L}{\partial \dot{x}})\delta \dot{x})_k\,dt = \sum_{k=1}^{w} p_k \sum_{i=1}^{N} \int_a^b \delta L_{ik}\,dt = \overline{\delta A}$$

where we used, for the particle $i$ with Hamiltonian $H_i$ and Lagrangian $L_i$, $F_{xi} = -\frac{\partial H_i}{\partial x_i} = \frac{\partial L_i}{\partial x_i}$,

$m_i \ddot{x} = \dot{p}_{xi} = \frac{\partial}{\partial t}(\frac{\partial L_i}{\partial \dot{x}_i})$, $\int_a^b \frac{\partial}{\partial t}(\delta x_j \frac{\partial L}{\partial \dot{x}})= (\delta x_j \frac{\partial L}{\partial \dot{x}})_a^b = 0$ due to the zero variation at *a* and *b*, and

the following definitions: $\overline{\delta A} = \sum_{k=1}^{w} p_k \delta A_k$ and $A_k = \int_a^b L_k\,dt$ (action calculated on the trajectory

$k$). Hence $\delta A_k = \sum_{i=1}^{N} \int_a^b \delta L_{ik}\,dt = \delta \int_a^b L_k\,dt$ with $L_k = \sum_{i=1}^{N} L_{ik}$ being the total Lagrangian of a system following the trajectory $k$.

We remember that Eq.(19) has been calculated as a sum of all the virtual works over a large but finite number of points during the period $\tau$ over the trajectories between *a* and *b*. As a matter of fact, at each moment of the motion, the principle of virtual work of d'Alembert applies, which implies that the virtual work of the ensemble at a given moment $t$ should vanish, just as in the case of equilibrium system with vanishing virtual work given by Eq.(9). We have

$$\delta W(t) = \sum_{k=1}^{w} p_k \delta W_k(t) = \sum_{k=1}^{w} p_k \sum_{i=1}^{N} [\overline{F}_i(t) - m_i \overline{a}_i(t)]_k \cdot \overline{\delta r}_{ik} = 0. \qquad (20)$$

This reasoning certainly leads to vanishing total virtual work between *a* and *b* since $\Delta W = \int_a^b \delta W(t) dt = 0$. By virtue of Eq.(19), we get the stochastic action principle of Eq.(1): $\overline{\delta A} = 0$. This is a derivation of SAP from the principle of virtual work.

6) Maximum path entropy

As mentioned above, Eq.(1) implies in fact an entropy variational approach. To see this, we calculate

$$\begin{aligned}\overline{\delta A} &= \sum_{j=1}^{w} p_j \delta A_j \\ &= \delta \sum_{j=1}^{w} p_j A_j - \sum_{j=1}^{w} \delta p_j A_j \\ &= \delta \overline{A} - \delta Q_{ab}\end{aligned} \qquad (21)$$

where $\overline{A} = \sum_{j=1}^{w} p_j A_j$ is the ensemble mean of action $A_j$, and $\delta Q_{ab}$ can be written as

$$\delta Q_{ab} = \delta \overline{A} - \overline{\delta A}. \qquad (22)$$

which can be considered as the definition of a path entropy, a measure of uncertainty of a the action on different possible trajectories. In mimicking the first law, $\delta Q_{ab} = \delta \overline{A} - \overline{\delta A}$ looks like a generalized "heat", a measure of the disorder of paths. If we introduce an 'inverse temperature' $\eta$ such that

$$\delta Q_{ab} = \frac{\delta S_{ab}}{\eta}, \qquad (23)$$

Then from Eq.(1) and Eq.(21), we get

$$\delta(S_{ab} - \eta \overline{A}) = 0. \qquad (24)$$

This is a variational calculus for a nonequilibrium random dynamics which optimizes the quantity $(Q_{ab} - \overline{A})$ and $(S_{ab} - \eta \overline{A})$. If the normalization condition is added as a constraint, Eq.(24) becomes :

$$\delta[S_{ab} - \eta \sum_j p_j A_j + \alpha \sum_j p_j] = 0 \qquad (25)$$

which is the maxent applied to path entropy with two Lagrange multipliers $\alpha$ and $\eta$, an approach originally proposed and investigated in the references [11][12]. In these previous works, it was shown that, is the path entropy takes the Shannon form, Eq.(25) yields an exponential path probability distribution of action. It was also shown in [12] that if the system is distributed in position space in the Gaussian way, the path probability is necessarily exponential and the path entropy defined by Eq.(23) is necessarily of Shannon form. This means that if, in a diffusion problem, the particles distribution is different, path entropy may take different forms. This is investigated in detailed way in reference [20].

## 7) Concluding remarks

After recapitulating an application of virtual work principle to equilibrium system in order to justify maxent with thermodynamic entropy, we presented an extension of this principle to thermodynamic system out of equilibrium in order to justify a least action principle $\overline{\delta A} = 0$ for the stochastic dynamics (SAP) or diffusion problem. This is carried out for Hamiltonian systems. One of the conclusions is that, in random mechanical system, the maximum thermodynamic entropy for equilibrium system or maximum path entropy for nonequilibrium system are the consequences of vanishing virtual work of the microscopic forces on every element of the system. Another conclusion of this work is that maxent in physics is not necessarily an inference method. It is a law of physics due to its tight correlation with a fundamental principle of physics.

# References

[1] P.L.M. de Maupertuis, *Essai de cosmologie* (Amsterdam, 1750) ; *Accord de différentes lois de la nature qui avaient jusqu'ici paru incompatibles.* (1744), Mém. As. Sc. Paris p. 417; *Le lois de mouvement et du repos, déduites d'un principe de métaphysique.* (1746) Mém. Ac. Berlin, p. 267

[2] V.I. Arnold, Mathematical methods of classical mechanics, second edition, Springer-Verlag, New York, 1989, p243

[3] L. Onsager and S. Machlup, Fluctuations and irreversible processes, *Phys. Rev.*, 91,1505(1953); L. Onsager, Reciprocal relations in irreversible processes I., *Phys. Rev.* 37, 405(1931)

[4] L. De Broglie, La thermodynamique de la particule isolée, Gauthier-Villars éditeur, Paris, 1964

[5] M.I. Freidlin and A.D. Wentzell, Random perturbation of dynamical systems, Springer-Verlag, New York, 1984

[6] G.L. Eyink, Action principle in nonequilibrium statistical dynamics, *Phys. Rev. E*, 54,3419(1996)

[7] F. Guerra and L. M. Morato, Quantization of dynamical systems and stochastic control theory, *Phys. Rev. D*, 27, 1774(1983)

[8] F. M. Pavon, Hamilton's principle in stochastic mechanics, *J. Math. Phys.*, 36, 6774(1995)

[9] R.P. Feynman and A.R. Hibbs, Quantum mechanics and path integrals, McGraw-Hill Publishing Company, New York, 1965

[10] S. Weinberg, Quantum field theory, vol.II, Cambridge University Press, Cambridge, 1996 (chapter 23: extended field configurations in particle physics and treatments of instantons)

[11] Q.A. Wang, Maximum path information and the principle of least action for chaotic system, *Chaos, Solitons & Fractals*, 23 (2004) 1253; Non quantum uncertainty relations of stochastic dynamics, *Chaos, Solitons & Fractals*, 26,1045(2005); Maximum entropy change and least action principle for

nonequilibrium systems, Astrophysics and Space Sciences, <u>305</u> (2006)273

[12] Q. A. Wang, F. Tsobnang, S. Bangoup, F. Dzangue, A. Jeatsa and A. Le Méhauté, Reformulation of a stochastic action principle for irregular dynamics, to appear in *Chaos, Solitons & Fractals*, (2007); arXiv:0704.0880

[13] E.T. Jaynes, The evolution of Carnot's principle, The opening talk at the EMBO Workshop on Maximum Entropy Methods in x-ray crystallographic and biological macromolecule structure determination, Orsay, France, April 24-28, 1984; Gibbs vs Boltzmann entropies, *American Journal of Physics*, <u>33</u>,391(1965) ; Where do we go from here? in *Maximum entropy and Bayesian methods in inverse problems*, pp.21-58, eddited by C. Ray Smith and W.T. Grandy Jr., D. Reidel, Publishing Company (1985)

[14] L.M. Martyushev and V.D. Seleznev, Maximum entropy production principle in physics, chemistry and biology, *Physics Reports*, <u>426</u>, 1-45 (2006)

[15] Jos Uffink, Can the maximum entropy principle be explained as a consistency requirement, *Studies in History and Philosophy of Modern Physics*, **26B** (1995): 223-261

[16] J.L. Lagrange,  Mécanique analytique, Blanchard, reprint , Paris  (1965)  (Also: Oeuvres, Vol. 11.)

[17] J. D'Alembert,  Traité de dynamique, Editions Jacques Cabay , Sceaux  (1990)

[18] C.G.Gray, G.Karl, V.A.Novikov, Progress in Classical and Quantum Variational Principles, Reports on Progress in Physics (2004), arXiv: physics/0312071

[19] Qiuping A. Wang, Seeing maximum entropy from the principle of virtual work, arXiv:0704.1076

[20] Q.A. Wang, Probability distribution and entropy as a measure of uncertainty, arXiv:cond-mat/0612076